\documentclass[sigconf]{acmart}
\AtBeginDocument{%
  }
\copyrightyear{2026}
\acmYear{2026}
\setcopyright{cc}
\setcctype{by-nc-nd}
\acmConference[ICMR '26]{International Conference on Multimedia Retrieval}{June 16--19, 2026}{Amsterdam, Netherlands}
\acmBooktitle{International Conference on Multimedia Retrieval (ICMR '26), June 16--19, 2026, Amsterdam, Netherlands}
\acmDOI{10.1145/3805622.3812606}
\acmISBN{979-8-4007-2617-0/2026/06}

\begin{document}

\title{Contestable Multi-Agent Debate with Arena-based Argumentative Computation for Multimedia Verification}


\author{Truong Thanh Hung Nguyen}
\authornote{Both authors contributed equally to this research.}
\orcid{0000-0002-6750-9536}
\affiliation{%
  \institution{University of New Brunswick}
  \city{Fredericton}
  \state{New Brunswick}
  \country{Canada}}
\email{hung.ntt@unb.ca}

\author{Vo Thanh Khang Nguyen}
\authornotemark[1]
\orcid{0009-0009-1633-7787}
\affiliation{%
  \institution{FPT Software}
  \city{Quy Nhon}
  \country{Vietnam}
}
\email{khangnvt1@fpt.com}

\author{Hoang-Loc Cao}
\orcid{0009-0009-5640-3144}
\affiliation{%
  \institution{University of Science, VNU-HCM}
  \city{Ho Chi Minh City}
  \country{Vietnam}}
\email{chloc22@clc.fitus.edu.vn}

\author{Phuc Ho}
\orcid{0009-0007-5842-2504}
\affiliation{%
  \institution{University of Science, VNU-HCM}
  \city{Ho Chi Minh City}
  \country{Vietnam}}
\email{hdphuc22@clc.fitus.edu.vn}

\author{Van Pham}
\orcid{0009-0004-5438-895X}
\affiliation{%
  \institution{University of Science, VNU-HCM}
  \city{Ho Chi Minh City}
  \country{Vietnam}}
\email{pavan22@clc.fitus.edu.vn}

\author{Hung Cao}
\orcid{0000-0002-0788-4377}
\affiliation{%
  \institution{University of New Brunswick}
  \city{Fredericton}
  \state{New Brunswick}
  \country{Canada}}
\email{hcao3@unb.ca}

\renewcommand{\shortauthors}{Hung Nguyen et al.}
\begin{teaserfigure}
    \centering
    \includegraphics[width=.9\linewidth]{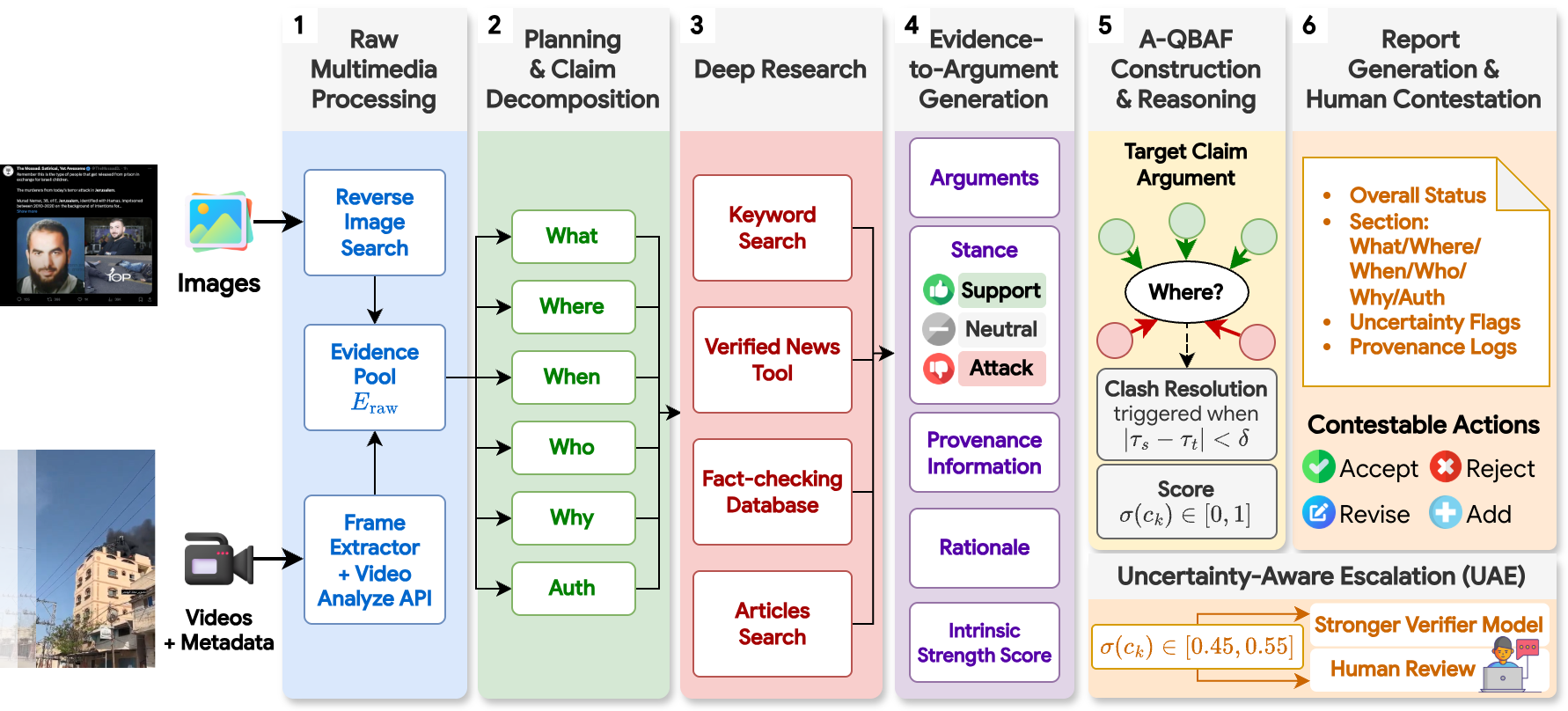}
    \caption{Architecture of Contestable Multi-Agent Debate with A-QBAF for Multimedia Verification.}
    \label{fig:arch}
\end{teaserfigure}

\begin{abstract}
Multimedia verification requires not only accurate conclusions but also transparent and contestable reasoning. We propose a contestable multi-agent framework that integrates multimodal large language models, external verification tools, and arena-based quantitative bipolar argumentation (A-QBAF) as a submission to the \textit{ICMR 2026 Grand Challenge on Multimedia Verification}. Our method decomposes each case into claim-centered sections, retrieves targeted evidence, and converts evidence into structured support and attack arguments with provenance and strength scores. These arguments are resolved through small local argument graphs with selective clash resolution and uncertainty-aware escalation. The resulting system generates section-wise verification reports that are transparent, editable, and computationally practical for real-world multimedia verification. Our implementation is public at: \url{https://github.com/Analytics-Everywhere-Lab/MV2026_the_liems}.
\end{abstract}





\maketitle

\section{Introduction and Related Works}
Multimedia verification is increasingly important because online misinformation is often conveyed through images and videos rather than solely through text \cite{dang2024overview,acmmm25-grand,dangNguyenMV2026}. Prior work has focused on two main problems: detecting manipulated or synthetic media such as deepfakes \cite{tolosana2020deepfakes}, and identifying authentic but miscontextualized content, often called cheapfakes or out-of-context misuse \cite{rossler2019faceforensics,aneja2022acm,guan2025nist}. As a result, research has expanded from classical media forensics to evidence-based verification that combines visual analysis, metadata inspection, reverse image search, fact-checking resources, and source credibility assessment \cite{woo2022advanced,moholdt2023detecting,pham2023detecting,seo2024multi}.

Recent studies increasingly use multimodal large language models (MLLMs) and hybrid vision-language systems to jointly reason over images, videos, and text \cite{tran2022textual,la2022multimodal,nguyen2023leveraging,nguyen2023multi}. These systems can also incorporate external tools such as reverse image search, metadata analysis, knowledge graphs, and fact-checking databases \cite{ganti2022novel,meredith2024osint,dao2023leveraging,nguyen2025multimodal,nguyen2025robust}. More recently, agent-based approaches have decomposed verification into multi-step workflows in which LLMs coordinate retrieval, tool use, and evidence synthesis \cite{nguyen2024hybrid,le2025multimedia}.

However, most existing approaches still emphasize prediction accuracy or final evidence synthesis rather than \emph{contestability} \cite{nguyen_motion2meaning_2025,nguyen2026heart2mind,alfrink2023contestable,ploug_four_2020}. In practice, they often compress heterogeneous evidence into a single conclusion, making it difficult for users to inspect supporting and attacking evidence, challenge intermediate reasoning, or revise conclusions when evidence is incomplete or disputed \cite{freedman_argumentative_2025}. This limitation is especially important in high-stakes multimedia verification, where conflicting signals, uncertain provenance, and interpretive claims such as \textit{why} a post was shared are common.

To address this gap, we propose a \emph{contestable multimedia verification} framework that combines MLLMs, external verification tools, and an arena-based quantitative bipolar argumentation framework (A-QBAF) \cite{cao2026adaptive}. Instead of directly summarizing evidence into a final answer, our method decomposes verification into claim-centered questions and converts evidence into structured support and attack arguments \cite{zhuargrag}. These arguments are organized into small local graphs, enabling transparent reasoning, efficient conflict handling, and human-editable contestation. The framework also includes uncertainty-aware escalation so that ambiguous cases can be deferred to a stronger verifier or human review. Our goal is to produce verification reports that are evidence-grounded, transparent, editable, and practical for real-world multimedia analysis.
Our contributions are as follows:
\begin{enumerate}
    \item We propose a \textbf{contestable multi-agent multimedia verification} framework that integrates MLLMs, external verification tools, and claim-centered A-QBAF reasoning to assess both authenticity and contextual integrity.
    \item We introduce an \textbf{evidence-to-argument formulation} that converts retrieved multimodal evidence into structured support and attack arguments with provenance, rationale, and strength scores, enabling users to inspect, accept, reject, edit, or add arguments during verification.
    \item We design an \textbf{efficient reasoning pipeline} based on small claim graphs, top-$k$ evidence selection, sparse relation construction, selective clash resolution, and uncertainty-aware escalation, producing structured reports that remain transparent and contestable without excessive computation.
\end{enumerate}

\section{Proposed Method}
\label{sec:method}

We propose a contestable multimedia verification framework that combines MLLMs, external verification tools, and the A-QBAF \cite{cao2026adaptive}. The main goal of our method is to verify both the authenticity and the context of multimedia content while keeping the reasoning process transparent, editable, and computationally efficient.

Our system contains six stages: (1) raw multimedia processing, (2) planning and claim decomposition, (3) section-wise deep research, (4) evidence-to-argument conversion, (5) A-QBAF reasoning with selective conflict resolution, (6) report generation and human contestation with uncertainty-aware escalation (UAE). 

Our system replaces direct evidence synthesis with a lightweight argumentation layer. Retrieved evidence is first converted into structured arguments, then organized into support-attack relations around six target claims: \emph{what}, \emph{where}, \emph{when}, \emph{who}, \emph{why}, and \emph{authenticity}. 
The framework is designed for practical multimedia verification: it handles both images and videos, integrates evidence from multiple tools and sources, identifies contradictions, and produces a structured report rather than only a binary label. To remain efficient, we use a \emph{section-wise micro-graph} design, where each claim is verified with a small local argument graph instead of a single global graph. This improves traceability and makes the final decision easier to inspect and contest.

\subsection{Raw Multimedia Processing}

The first stage prepares the multimedia input for later verification. Given a case package $\mathcal{I}$, the system receives:
\begin{equation}
\mathcal{I} = \{m, d, u\},
\end{equation}
where $m$ is the primary media content (image or video), $d$ is the available metadata and textual context, and $u$ represents optional external clues such as source links, captions, or fact-checking notes.

For video input, the system uses an MLLM analyzer to generate frame-level descriptions, extract visible text, identify key objects and events, and collect technical metadata. A keyframe extractor selects the most informative frames to reduce redundancy and later search cost. For image input, the system performs reverse image search and collects visually related web pages, reposts, and possible original sources.
This stage outputs a normalized evidence pool:
\begin{equation}
E_{raw} = \{e_1, e_2, \dots, e_n\},
\end{equation}
where each evidence may come from a frame, metadata field, OCR result, ASR transcript, reverse search result, or retrieved article.

\subsection{Planning and Claim Decomposition}
The planner agent coordinates the verification workflow. Instead of treating the case as one single prediction problem, the planner decomposes it into a set of claim-centered verification questions:
\begin{equation}
C = \{c^{what}, c^{where}, c^{when}, c^{who}, c^{why}, c^{auth}\}.
\end{equation}

These claims are defined as follows:
\begin{itemize}
    \item $c^{what}$: what source event or situation is shown,
    \item $c^{where}$: where the content was recorded,
    \item $c^{when}$: when the event happened,
    \item $c^{who}$: which people, groups, or sources are involved,
    \item $c^{why}$: why content was posted or what narrative it supports,
    \item $c^{auth}$: whether the media is authentic, edited, synthetic, or recaptured.
\end{itemize}

This decomposition is important for two reasons. First, some claims are more objective than others. For example, \emph{where} and \emph{when} can often be grounded by landmarks, timestamps, or metadata, while \emph{why} is usually more interpretive and requires stronger explanation. Second, not every case needs the same amount of reasoning. By separating claims, the system can apply more computation only to the difficult parts.

\subsection{Deep Research}

After claim decomposition, the deep researcher agent gathers targeted evidence for each claim. For a claim $c_k \in C$, the agent retrieves a small relevant evidence subset:
\begin{equation}
E_k = \text{Top}^k(E_\text{raw}, c_k, k),
\end{equation}
where $\text{Top}^k$ selects the most relevant evidence items for the target claim. In practice, we keep $k$ small to control cost.

The deep researcher combines three types of operations:
\begin{enumerate}
    \item \textbf{keyword-guided search}: generate search queries from the media content and textual clues;
    \item \textbf{tool-assisted verification}: use reverse image search, metadata inspection, OCR/ASR, and fact-checking databases;
    \item \textbf{source analysis}: inspect article credibility, publication time, repost chains, and cross-source consistency.
\end{enumerate}

This stage is especially important for claims such as \emph{where}, \emph{when}, \emph{who}, and \emph{why}. For example, a verified news source may provide explicit location and date information, while a video frame may contain overlays or landmarks that either support or contradict that source. Our method keeps all such signals instead of collapsing them too early into a free-form summary.

\subsection{Evidence-to-Argument Generation}

The main contribution of our method begins at this stage. We convert retrieved evidence into structured arguments. For each evidence item $e_i \in E_k$, the system creates an argument card:
\begin{equation}
a_i = \{t_i, s_i, p_i, r_i, \tau_i\},
\end{equation}
\begin{itemize}
    \item $t_i$ is the textual form of the argument,
    \item $s_i \in \{\texttt{support}, \texttt{attack}, \texttt{neutral}\}$ is its stance toward the claim,
    \item $p_i$ stores provenance information,
    \item $r_i$ is a short rationale explaining why the evidence matters,
    \item $\tau_i \in [0,1]$ is the intrinsic strength score.
\end{itemize}

Each argument is written in a simple, evidence-grounded form. For instance, in a \textit{when} claim, one argument may state that ``The post was published on \textit{May 7, 2025}, during the India–Pakistan escalation.''  While another may attack the same claim by noting that the event timestamps were mismatched, i.e., ``Pakistani press coverage of the crash appeared on \textit{April 16, 2025}, well before Operation Sindoor.''

\paragraph{Intrinsic Strength Attribution.}
We assign a base score to each argument using four lightweight criteria:
\begin{equation}
\tau_i = \lambda_1 q_i^{src} + \lambda_2 q_i^{corr} + \lambda_3 q_i^{mod} + \lambda_4 q_i^{rel},
\end{equation}
where $q_i^{src}$ is source reliability, $q_i^{corr}$ is cross-source corroboration, $q_i^{mod}$ is cross-modal consistency, $q_i^{rel}$ is claim relevance.
The weights satisfy $\sum_j \lambda_j = 1$. In our implementation, this score is intentionally simplified. We avoid a fully debate-based scoring process for every argument because it would be overly expensive for multimedia cases with many noisy observations.

A direct multi-agent debate over all evidence would increase computational cost and often lead to redundant information. Our formulation is cheaper because we first select a small evidence subset for each claim, then convert only those items into short argument cards. This retains structure without creating unnecessarily long reasoning traces.

\subsection{A-QBAF Construction and Reasoning}
For each claim $c_k$, we build one small A-QBAF~\cite{cao2026adaptive}:
\begin{equation}
\mathcal{G}_k = \langle X_k, R_k^-, R_k^+, \tau_k \rangle,
\end{equation}
where $X_k = \{c_k\} \cup A_k$ is the set of nodes, including the target claim and its arguments; $R_k^+$ is the set of support edges; $R_k^-$ is the set of attack edges; $\tau_k$ is the base strength function.
The target claim node is initialized with neutral strength $\tau(c_k) = 0.5$.

If argument $a_i$ supports the claim, then $(a_i, c_k) \in R_k^+$. If it attacks the claim, then $(a_i, c_k) \in R_k^-$. We also allow limited argument-to-argument edges when two arguments clearly reinforce or contradict each other. However, to keep the graph sparse, we only add these edges in obvious cases, such as:
\begin{itemize}
    \item two arguments based on the same visual cue but opposite interpretations,
    \item metadata directly contradicting a reported date or location,
    \item two independent trusted sources confirming the same fact.
\end{itemize}

This sparse graph design is important. Full pairwise relation prediction across all arguments is expensive and often unnecessary in multimedia verification. In many cases, the main structure is already captured by argument-to-claim edges.

\subsection{Report Generation and Human Contestation}
\subsubsection{Quantitative Reasoning and Conflict Calibration}\label{ss:conflict}
After graph construction, we propagate argument influence using quantitative bipolar reasoning. For a node $x \in X_k$, its total incoming energy is:
\begin{equation}
E(x) = \sum_{y \in Sup(x)} \sigma(y) - \sum_{z \in Att(x)} \sigma(z),
\end{equation}
where $\sigma(\cdot)$ is the propagated strength.

We use the quadratic energy function:
$h(x) = \frac{\max(x,0)^2}{1 + \max(x,0)^2}$.
The final equilibrium score of node $x$ is:
\begin{equation}
\sigma(x) = \tau(x) + (1-\tau(x))h(E(x)) - \tau(x)h(-E(x)).
\end{equation}

The final claim score $\sigma(c_k)$ reflects the balance between supporting and attacking evidence. A high score means the claim is strongly supported by the available evidence. A low score means the claim is weakened by contradictions. A score near $0.5$ means the evidence remains inconclusive.

\paragraph{Selective Clash Resolution (CR)}
A common problem in LLM-based reasoning is score saturation, in which both support and attack arguments appear equally strong. To address this, we added a lightweight CR step. For a support--attack pair $(a_s, a_t)$, if their base scores are close
$|\tau(a_s) - \tau(a_t)| < \delta,$
we trigger one additional comparison by a judge model. This judge decides which argument is stronger, given the claim and the case evidence.
The score update is:
$\Delta \tau(a) = \beta(2w - 1),$
where $w \in [0,1]$ is the win rate of the argument across clashes and $\beta$ is a small adjustment magnitude. We only apply CR to ambiguous sections that benefit most from conflict calibration. For easier sections, such as locations with strong landmark evidence, this extra step is usually skipped.

\subsubsection{Human Contestation and Editable Verification}

One important goal of our method is contestability. The user should not only read the final report, but also inspect and challenge the reasoning process. Therefore, for each claim, the system presents: (1) the claim statement; (2) the main support and attack arguments; (3) provenance links to frames, metadata, or articles; (4) both intrinsic and propagated scores; and (5) a short explanation of why each argument affects the conclusion.

Users may perform four contestable actions: (1) accept an argument, (2) reject an argument, (3) edit an argument, or (4) add a missing argument with evidence.
After any accepted edit, the graph is updated and all claim scores are recomputed. In this way, contestation changes the final result directly rather than appearing as post-hoc feedback.
This design is especially useful for difficult claims such as \emph{why}. In many real cases, the system cannot fully prove the motivation behind a post, but it can show which arguments support a likely interpretation and which arguments weaken it. This makes the result more honest and easier to audit.

\subsubsection{Uncertainty-Aware Escalation (UAE)}
However, not all cases warrant a definitive system-level decision. If a claim score remains close to the neutral region, the system should admit uncertainty and request a stronger review. For claim $c_k$, if: $\sigma(c_k) \in [0.45, 0.55],$ we mark the section as uncertain.
For uncertain claims, we use one of two escalation options:
(1) invoke a stronger verifier model for one final judgment, or (2) defer the claim to human review.
This mechanism prevents the system from giving overconfident conclusions when the evidence is balanced or incomplete. It is particularly important for authenticity judgments and narrative interpretation, where both false positives/negatives are costly.

\subsection{Final Report Generation}

The final report combines section scores into a structured verification document. Instead of a single binary label, the report includes:
(1) overall verification status, (2) section-wise findings for \emph{what}, \emph{where}, \emph{when}, \emph{who}, \emph{why}, and \emph{authenticity}, (3) key supporting and attacking evidence, (4) uncertainty notes for inconclusive claims, (5) provenance records and contestation logs.
This report format matches the practical needs of multimedia fact-checking. It explains not only what the system concludes, but also why the conclusion was reached, which evidence supports it, which evidence challenges it, and which parts still remain uncertain.

Our method is designed to improve contestability without incurring high computational cost. It achieves this through several efficiency choices, i.e., claim decomposition into small local graphs, top-$k$ evidence selection for each claim, sparse relation construction, selective CR only for near-tied arguments, and escalation only for uncertain cases, where the input evidence is multimodal, noisy, and often large-scale. Our method adds argument structure only where it is most useful, rather than turning the whole pipeline into a heavy debate system.

\section{Demonstrative Results and Conclusion}
To illustrate the performance of our framework, we present an argumentative reasoning example using sample ID01 from the \textit{ICMR 2026 Grand Challenge on Multimedia Verification} \cite{dangNguyenMV2026} validation set. In this experiment, Gemini 3.1 Pro, GPT 5.4, and Claude Opus 4.6 were used as debaters to generate evidence-grounded arguments, which were then organized and resolved through the proposed reasoning A-QBAF.
\begin{figure}[h]
    \centering
    \includegraphics[width=0.8\linewidth]{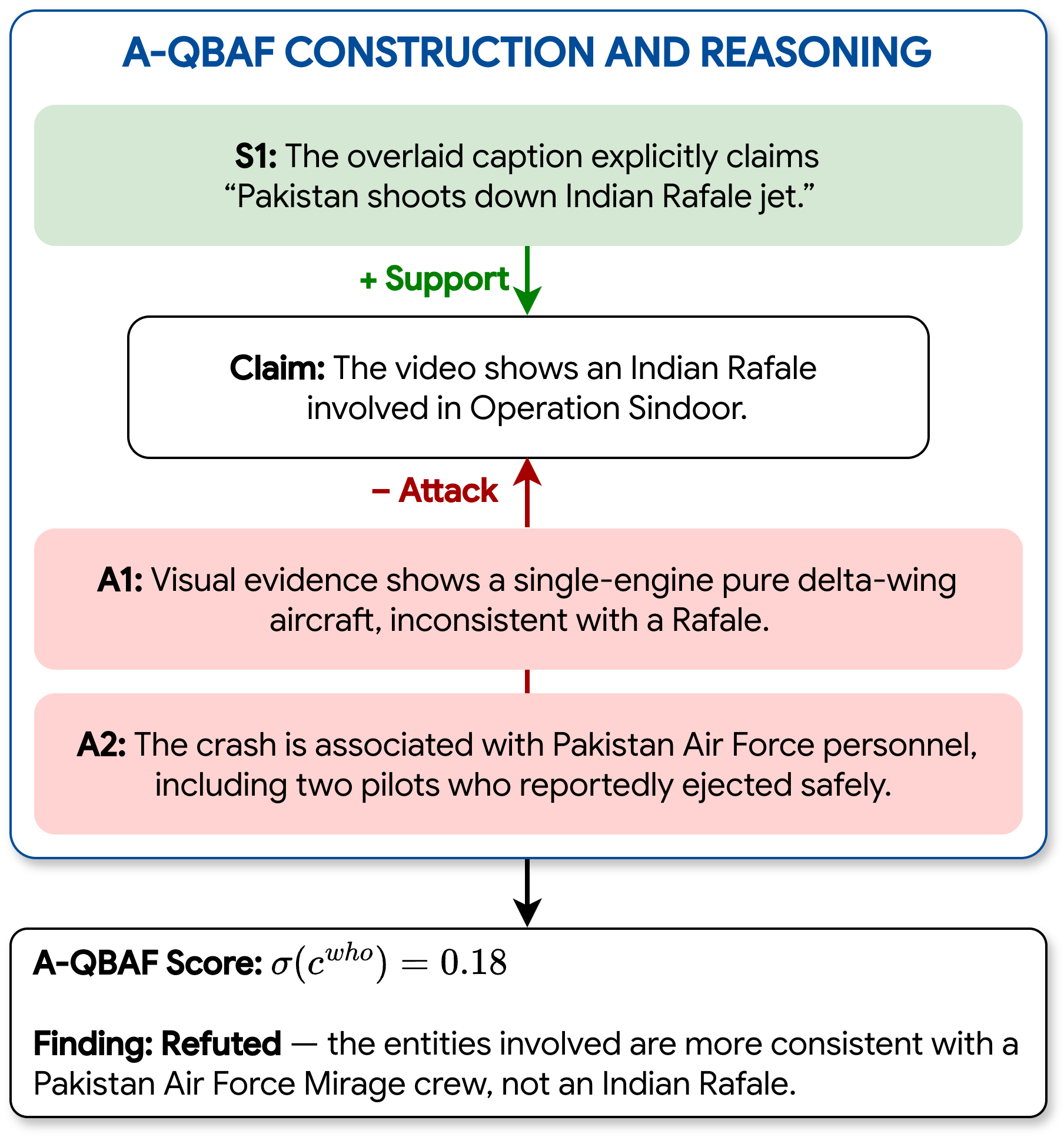}
    \caption{The final section-wise conclusion for \textit{who} (ID01 Validation) via argumentative A-QBAF computation.}
    \label{fig:sample}
\end{figure}

Fig.~\ref{fig:sample} shows the final reasoning outcome for the \textit{who} claim. The target claim states that the video shows an Indian Rafale involved in Operation Sindoor. One supporting argument is derived from the overlaid caption. However, two attacking arguments receive stronger overall influence in the graph. After A-QBAF reasoning, the final claim score is \(\sigma(c^{who})=0.18\), indicating that the claim is strongly weakened by the available evidence. The resulting section-level finding is therefore \textit{refuted}.

This example shows the advantages of our framework, i.e., transparent support-attack reasoning, human contestability through editable arguments and score recomputation, and practical efficiency through a small local graph. It also highlights uncertainty calibration. Since the final score lies well outside the uncertainty band defined in Sec~\ref{ss:conflict}, this claim does not require escalation to a stronger verifier or human review. More broadly, the same claim-centered process can be applied to \textit{what}, \textit{where}, \textit{when}, \textit{who}, \textit{why}, and \textit{authenticity}, producing a section-wise report that is more informative than a single binary label.

In conclusion, we propose a contestable multimedia verification framework by combining evidence retrieval, evidence-to-argument conversion, A-QBAF reasoning, selective CR, and UAE. The framework produces verification reports that are transparent, editable, and practical for high-stakes multimedia analysis.

\begin{acks}
This work was supported by NSERC Discovery Grant No RGPIN-2025-04478 and NSERC Discovery Supplement Award No DGECR-2025-00129.
\end{acks}
\balance{
    \bibliographystyle{ACM-Reference-Format}
    \bibliography{acmart}
}
\end{document}